\documentclass{iopart}
\usepackage{graphicx}
\usepackage{epsfig}
\usepackage{psfrag}
\usepackage{amssymb}
\newcommand{\be}{\begin{equation}}
\newcommand{\ee}{\end{equation}}

\newcommand{\rd}{\,{\rm d}}
\newcommand{\avec}{\mbox{\boldmath$a$}}

\newcommand{\dvec}{\mbox{\boldmath$d$}}
\newcommand{\rvec}{\mbox{\boldmath$r$}}
\newcommand{\tvec}{\mbox{\boldmath$t$}}

\begin{document}
\title{LISA source confusion: identification and characterization of signals}
%
\author{Richard Umst\"atter$^1$,
Nelson Christensen$^2$, Martin
Hendry$^3$, Renate
Meyer$^1$, Vimal
Simha$^3$, John
Veitch$^3$, Sarah
Vigeland$^2$ and Graham
Woan$^3$ }
\address{$^1$Department of Statistics, University of Auckland,
Auckland, New Zealand\\
$^2$Physics and Astronomy, Carleton College,
Northfield, MN 55057, USA\\
$^3$Department of Physics and Astronomy, University of Glasgow,
Glasgow G12\,8QQ, UK}
\ead{richard@stat.auckland.ac.nz, nchriste@carleton.edu,
martin@astro.gla.ac.uk, meyer@stat.auckland.ac.nz,
vimal\_simha@hotmail.com, jveitch@astro.gla.ac.uk,
vigelans@carleton.edu, graham@astro.gla.ac.uk}
\date{\today}
\begin{abstract}
The Laser Interferometer Space Antenna  (LISA) is expected to
detect gravitational radiation from a large number of compact
binary systems. We present a method by which these signals can be
identified and have their parameters estimated. Our approach uses
Bayesian inference, specifically the application of a Markov chain
Monte Carlo method. The simulation study that we present here
considers a large number of sinusoidal signals in noise, and our
method estimates the number of periodic signals present in the
data, the parameters for these signals and the noise level. The
method is significantly better than classical spectral techniques
at performing these tasks and does not use stopping criteria for
estimating the number of signals present.
\end{abstract}
\pacs{04.80.Nn, 02.70.Lq, 06.20.Dq}
%
\maketitle
\section{Introduction}
LISA~\cite{LISA} is expected to detect a very large number of
signals from compact binaries in the $10^{-2}$\,mHz to $100$\,mHz
band, making signal identification very difficult. Tens of
thousands of signals could be present in the data with significant
signal-to-noise ratios. In the 0.1\,mHz to 3\,mHz band there will
be numerous signals from white dwarf binaries. Sources above
5\,mHz should be resolvable, however below 1\,mHz there will be
\emph{source confusion}. In the 1\,mHz to 5\,mHz band we expect as
many as $10^5$ potential sources~\cite{cutler,bena,cornish}
resulting in an astoundingly difficult data analysis problem. We
direct the reader to Barack \& Cutler \cite{cutler} and Nelemans
et al.\ \cite{nele01} for an in-depth description of the
population of binary systems in the LISA operating band, and how
LISA's performance is influenced by them.

The goal of this paper is to introduce the LISA data analysis
community to a new approach for identifying and characterizing
these numerous signals. We apply Bayesian Markov chain Monte Carlo
(MCMC) methods to a simplified problem that will serve as an
example of the technique. MCMC methods are a numerical means of
parameter estimation, and are especially useful when there are a
large number of parameters~\cite{gilks96}. We have already applied
MCMC methods to other gravitational radiation parameter estimation
problems; for example, we have used a Metropolis-Hastings (MH)
algorithm \cite{metr53,hastings70} for estimating astrophysical
parameters for gravitational wave signals from coalescing compact
binary systems \cite{insp}, and pulsars \cite{pul1,pul2}. We
believe that MCMC methods could provide an effective means for
identifying sources in LISA data. We summarize our reversible jump
MCMC technique in this paper. A more detailed and comprehensive
description can be found in~\cite{LISAMCMC1}.

Here we present a summary of our study of some simple simulated
data, comprising a number of sinusoidal signals embedded in noise.
Our reversible jump MCMC algorithm infers the parameters for each
(sufficiently large) sinusoidal signal, the magnitude of the noise
and the number of signals present. In our approach we solve both
the detection and parameter estimation problems without the need
for evaluating formal model selection criteria. The method does
not require a stopping criterion for determining the number of
signals and produces results which compare very favorably with
classical spectral techniques. A Bayesian analysis naturally
encompasses Occam's Razor and a preference for a simpler model {\cite{Jaynes03}}. In
addition, our MCMC method is better than a classical periodogram
at resolving signals that are very close in frequency, and we
provide an explanation of how to identify these signals.

The method that we present here is not a \emph{source subtraction}
method \cite{cornish2}. Signals that are sufficiently strong will
be identified with a quoted confidence, and sources that are weak
will simply contribute to the noise, the level of which we also
estimate. We show that the noise level estimate from our method
depends (as it should) on the inherent detector noise level, and
also the presence of unidentified signals. A benefit of using MCMC
methods is that computation time does not show an exponential increase
with the number of parameters
\cite{gilks96}.

The problem of identifying an unknown number of sinusoids is
neither new nor simple \cite{Rich,Andrieu}. Previous studies have
looked for a handful of unknown signals, here we show results for
100 signals. MCMC methods are robust and dynamic, and we believe
that ultimately it will be possible to use them with LISA data to
estimate the parameters of all modelled sources types. In the
future we will make the model more complex, taking into account
the orbit of the LISA spacecraft and binary source evolution.

\section{Occam factors}
\label{model} One can approach the problem of identifying and
enumerating sinusoidal signals in noise from a number of different
directions, but one thing is clear. Discrete noisy data can be
fitted \emph{exactly} if one uses a sufficient number of
components -- the result is simply the discrete Fourier transform
of the data. Classically, we proceed by estimating the noise floor
of the spectrum and identify a threshold spectral power that
divides the components between signals and noise.  In this way we
prevent the model from overfitting the noise.  In an iterative
fitting procedure, this is achieved by halting when the statistics
of the residuals fit the noise model well. One very attractive,
and well known, feature of Bayesian inference is that these ideas
are within the fabric of the method. Indeed they are such a basic
property of logical inference that there is no need to refer to
ideas such as `overfitting' at all.  More generally, the method
discourages us from using models that have more degrees of freedom
than are necessary for the problem in hand.

One can see how this works in a simple example: Take two data
models, $\mathcal{M}_1$ and $\mathcal{M}_2$ constrained by a
single datum, $d$. $\mathcal{M}_1$ has one parameter, $a_1$, to
describe the datum, whereas $\mathcal{M}_2$ uses the sum of two
parameters, $s=a_1+a_2$ to describe the same datum. Which model is
better? Here we have no noise and no random variables, so this is
not a problem for orthodox statistics. However,  if the datum is
equally consistent with both models, we would clearly prefer the
simpler model $\mathcal{M}_1$ in favour of $\mathcal{M}_2$.

\begin{figure}[hbt]
\centerline{\includegraphics[width=7cm]{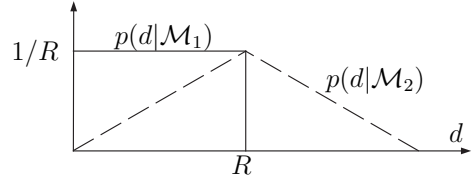} } \caption{The
evidences for models $\mathcal{M}_1$ (solid line, one parameter
fit) and $\mathcal{M}_2$ (dashed line, two parameter fit) as a
function of the datum $d$. Note that the evidence ratio always
favours the simpler model $\mathcal{M}_1$ when the datum is
equally consistent with either.} \label{fig_1}
\end{figure}

Within the Bayesian framework we consider the odds ratio of the
two models:
\begin{equation}
\mathcal{O}_{12}=\frac{p(\mathcal{M}_1|d) }{p(\mathcal{M}_2|d)}
=\frac{p(\mathcal{M}_1)}
{p(\mathcal{M}_2)}\frac{p(d|\mathcal{M}_1)}{p(d|\mathcal{M}_2)}.
\end{equation}
We will set  $p(\mathcal{M}_1)/p(\mathcal{M}_2)$ to unity, as we
have no prior preference for either model, and take the priors for
$a_1$ and $a_2$ to be each uniform in the range $0\rightarrow R$,
making the prior for $s$ the convolution of two of these. The
functional priors for $a_1$ under $\mathcal{M}_1$ and $s$ under
$\mathcal{M}_2$ are therefore
\begin{equation}
\fl
p(a_1|\mathcal{M}_1) = \frac{1}{R};\qquad p(s|\mathcal{M}_2)
=\cases{
s/R^2 & \kern -2em for $0<s<R$\\
2/R-s/R^2  & \kern -2em for $R<s<2R$. }
\end{equation}
The probability of the data, given either model, is simply a Dirac
delta function centred on the value of the datum, so we can
calculate the \emph{evidences} $p(d|\mathcal{M}_1)$ and
$p(d|\mathcal{M}_2)$ by marginalising over the allowed parameter
values:
\begin{eqnarray}
\fl p(d|\mathcal{M}_1)=\int_0^R\frac{1}{R}\delta(d-a_1)\,\rd a_1 =
\cases{
1/R & \kern -2em for $0<d<R$\\
0 & \kern -2em otherwise,
}\\
\fl p(d|\mathcal{M}_2) =
\int_0^{2R}p(s|\mathcal{M}_2)\delta(d-s)\,\rd s = \cases{
d/R^2 & \kern -2em for $0<d<R$\\
2/R-d/R^2  & \kern -2em for $R<d<2R$\\
0     & \kern -2em otherwise.}
\end{eqnarray}
as shown in Fig.~\ref{fig_1}.
If the datum lies in the range $0<d<R$, the odds ratio is
$(1/R)/(d/R^2)=R/d$, so that $\mathcal{M}_1$ will always favoured
over $\mathcal{M}_2$ in that range.  If $d=R$ the odds ratio is
unity, and if $R<d<2R$ the only possible model is $\mathcal{M}_2$.

This demonstrates why the simpler model is favoured even when the
datum is equally consistent with both: $\mathcal{M}_2$ has more
flexibility than is necessary to explain the datum and so
penalizes itself by spreading its evidence more thinly.

Although in this example we consider a probability ratio to
determine our favoured model, when more than two models are
available we can consider the model choice to be a parameter
itself, and determine its marginal posterior probability in the
usual way.

A slightly more pertinent, though still highly restricted, example
would be a data set $\{d_k\}$ that consists of observations of the
sum of $m$ sinusoids of the form $A_i\sin(2\pi f_i t)$ at times
$t=t_k$ and Gaussian noise with variance $\sigma^2$. We also know
that $0\le m\le5$, and $A_i$ and $f_i$ can only take the discrete
values $A_i\in \{1,\ldots,5\}$ and $f_i\in \{0.01,
0.02,\ldots,0.05\}$.

For this discrete problem, assuming uniform priors on $m$, $A_i$
and $f_i$ for each sinusoid, we can identify the probability of
any particular $m$, given the data, irrespective of the other
signal parameters:
\begin{equation}
p(m|\{d_k\}) \propto
\frac{1}{25^m}\sum_{A_1,f_1}\ldots\sum_{A_M,f_M}\exp(-\chi^2/2)
\label{discrete}
\end{equation}
where
\begin{equation}
\chi^2 = \frac{1}{\sigma^2}\sum_k\left[d_k-\sum_{i=1}^m
A_i\sin(2\pi f_i t_k)\right]^2.
\end{equation}
Here it is the factor of $25^m$, originating from the $m$
normalised priors for the model parameters, that offsets the
increasingly good $\chi^2$ fit that might come from large values
of $m$ and provides our Occam factor.

This discrete problem can be solved by directly marginalising over
the nuisance amplitude and frequency parameters.  However problems
with more and/or with continuous parameters are not
approachable using such direct methods, and must be tackled in
another way.
\section{Parameter estimation}
\label{MH} \label{TMHA} We consider the continuous case as a
signal consisting of $m$ superimposed sinusoids where $m$ is
unknown. Therefore we confine our attention to a set of models $\{
\mathcal{M}_m: m \in \{1,\cdots,M\} \}$ where $M$ is the maximum
allowable number of sinusoids. Let $\dvec=[d_1,\cdots,d_N]$ be a
vector of $N$ samples recorded at times $\tvec=[t_1,\cdots,t_N]$.
Model $\mathcal{M}_m$ assumes that the observed data are composed
of a signal plus noise: $d_j=\mathrm{f}_m(t_j,\avec_m)+
\epsilon_j$, for $j=1\ldots N$, where the noise terms $\epsilon_j$
are assumed to be i.i.d.\ $N(0,\sigma_m^2)$ random variables. The
signal of model $ \mathcal{M}_m$ is assumed to be of the form
\begin{equation}
\mathrm{f}_m(t_j,\avec_m)=\sum_{i=1}^{m}{\left[ A_i^{(m)} \cos(2
\pi f_i^{(m)} t_j)+B_i^{(m)} \sin(2 \pi f_i^{(m)} t_j) \right]}.
\label{signal}
\end{equation}
Model $\mathcal{M}_m$ is therefore characterized by a vector
\begin{equation}
\avec_m=[A_1^{(m)},B_1^{(m)},f_1^{(m)},\cdots,A_m^{(m)},B_m^{(m)},f_m^{(m)},\sigma^2_m]
\end{equation}
of $3m+1$ unknown parameters. The objective is  to find the model
$\mathcal{M}_m$ that best fits the data. To this end, we use a
Bayesian approach as in {\cite{Bretthorst88}}. The joint
probability of these data $\dvec$ given the parameter vector
$\avec_m$ and model $\mathcal{M}_m$ is
\begin{equation}
p(\dvec|m,\avec_m) \propto \frac{1}{\sigma^N_m} \exp \left\{ -
\frac{1}{2 \sigma^2_m} \sum_{j=1}^N \left[d_j -
\mathrm{f}_m(t_j,\avec_m)\right]^2 \right\}.
\end{equation}
We choose (now continuous) uniform priors for the amplitudes
$A_i^{(m)}$,$B_i^{(m)}$ and frequency $f_i^{(m)}$ with ranges
$A_i^{(m)} \in [-A_{\rm max},A_{\rm max}]$, $ B_i^{(m)}
\in[-B_{\rm max},B_{\rm max}]$ and $f_i^{(m)}  \in [0,0.5]$, respectively. Furthermore we use a uniform prior for $m$
over $\{1,\ldots,M\}$ and vague inverse Gamma priors for
$\sigma_m^2$.
 By applying Bayes' theorem, we obtain the posterior pdf
\begin{equation}
p(m,\avec_m|\dvec) = \frac{ p(m,\avec_m) p(\dvec|m,\avec_m)
}{p(\dvec)}, \label{posterior}
\end{equation}
where $p(\dvec)=\sum_{i=1}^M \int p(m,\avec_m) p(\dvec|m,\avec_m)
\,\rd\avec_m$. We use a sampling-based technique for posterior
inference via MCMC \cite{gilks96}. MCMC techniques only require
the unnormalised posterior $p(m,\avec_m|\dvec) \propto
p(m,\avec_m) p(\dvec|m,\avec_m)$ to simulate from
Eq.~(\ref{posterior}) in order to estimate the quantities of
interest. However, as the dynamic variable of the simulation does
not have fixed dimension, the classical MH techniques
\cite{metr53,hastings70} cannot be adopted when proposing
trans-dimensional moves between models where the model indicator
$m$ determines the dimension ($3m+1$) of the parameter vector
$\avec_m$. We therefore use the Reversible Jump Markov Chain Monte
Carlo (RJMCMC) algorithm \cite{Green95,Green03} for model
determination, as in \cite{Andrieu}. For transitions within the same
model, we use the
delayed rejection method \cite{Mira98,TierneyMira99} which yields
a better adaptation of
the proposals in different parts of the state space.

\subsection{The RJMCMC for model determination}
To sample from the joint posterior $p(m,\avec_m|\dvec)$ via MCMC,
we need to construct a Markov chain simulation with state space
$\displaystyle\cup_{m=1}^M \left({m}\times I\!\!R^{3m+1}\right)$. When
a new model is proposed we attempt a step between state spaces of
different dimensionality. Suppose that at the $n$th iteration of
the Markov chain we are in state $(k,\avec_k)$. If model
$\mathcal{M}_{k'}$ with parameter vector $\avec_{k'}'$ is
proposed, a reversible move has to be considered in order to
preserve the detailed balance equations of the Markov chain.
Therefore the dimensions of the models have to be matched by
involving a random vector $\rvec$ sampled from a proposal
distribution with pdf $q$, say, for proposing the  new parameters
$\avec_{k'}'=\textrm{t}(\avec_k,\rvec)$ where $\textrm{t}$ is a
suitable deterministic function of the current state and the
random numbers. Here we focus on transitions that either decrease
or increase models by one signal, i.e.\ $k'\in \{k-1,k+1\}$. We
use equal probabilities $p_{k\mapsto k'}=p_{{k' \mapsto k}}$ to
either move up or down in dimensionality. Without loss of
generality, we consider $k<k'$.

If  the transformation $\textrm{t}_{k\mapsto k'}$ from
$(\avec_k,\rvec)$ to $\avec_{k'}'$ and its inverse
$\textrm{t}_{k\mapsto k'}^{-1}=\textrm{t}_{k'\mapsto k}$ are both
differentiable, then reversibility is guaranteed by defining the
acceptance probability for increasing a model by one signal
according to \cite{Green95} by
\begin{equation}\label{eq:green}
\alpha_{k \mapsto k'}(\avec_{k'}'|\avec_k)= \min \left\{ 1, \frac{
p(\avec_{k'}',k') p(\dvec|\avec'_{k'},k')p_{k\mapsto k'}
}{p(\avec_k,k) p(\dvec|\avec_k,k) q(\rvec) p_{k'\mapsto k} }
\right\} \left| J_{k\mapsto k'} \right|
\end{equation}
where $|J_{k\mapsto k'}|=\left|\frac{\partial
\textrm{t}(\avec_k,\rvec)}{\partial(\avec_k,\rvec)}\right|$ is the
Jacobian determinant of this transformation and $q(\rvec)$ is the
proposal distribution.

In this context, two types of transformations, `split-and-merge'
and `birth-and-death', are obvious choices. In a `split-and-merge'
transition, the proposed parameter vector $\avec'_{k'}$ comprises
all $(k-1)$ subvectors of $\avec_k$ except a randomly chosen
subvector $\avec_{(i)}=(A_i^{(k)},B_i^{(k)},f_i^{(k)})'$ which is
replaced by two 3-dimensional subvectors,
$$\avec_{(i_1)}'=\left(A_{i_1}^{(k')},B_{i_1}^{(k')},f_{i_1}^{(k')}\right)
\quad {\rm and}\quad \avec_{(i_2)}'=\left(
A_{i_2}^{(k')},B_{i_2}^{(k')},f_{i_2}^{(k')}\right)$$ with roughly half
the amplitudes but about the same frequency as  $\avec_{(i)}$.

A three-dimensional Gaussian random vector (with mean
zero), $\rvec=(r_A,r_B,r_f)$, changes the current state
$\avec_{(i)}$ to the two resulting states $\avec_{(1i)}',
\avec_{(2i)}'$ through a linear transformation
\begin{eqnarray}
\fl \textrm{t}_{k\mapsto k'}(\avec_{(i)},\rvec)=\left(
\frac{1}{2}A_i^{(k)}\!+r_A, \frac{1}{2}B_i^{(k)}\!+r_B,
f_i^{(k)}\!+r_f, \frac{1}{2}A_i^{(k)}\!-r_A,
\frac{1}{2}B_i^{(k)}\!-r_B, f_i^{(k)}\!-r_f
\right)\nonumber\\
\fl \phantom{\textrm{t}_{k\mapsto k'}(\avec_{(i)},\rvec)}=
\left(A_{i_1}^{(k')},B_{i_1}^{(k')},f_{i_1}^{(k')},A_{i_2}^{(k')},B_{i_2}^{(k')},f_{i_2}^{(k')}\right).
\end{eqnarray}
By analogy,  the inverse transformation $\textrm{t}^{-1}_{k\mapsto
k'}:=\textrm{t}_{k'\mapsto k}$ accounts for the merger of two
signals. Note that the determinant of the Jacobian of the transformation
$\textrm{t}_{k\mapsto k'}$ is
$|J_{k\mapsto k'}|=2$, and that of its inverse is 1/2.

We use a multivariate normal distribution,
$N[\textbf{0},\textrm{diag}(\sigma_A^2,\sigma_B^2,\sigma_f^2)]$,
for the proposal distribution $q(\rvec)$.
Care has to be taken in choosing suitable values for the proposal
variances to achieve reasonable acceptance probabilities in  Eq.~(\ref{eq:green}).

The second `birth-and-death' transformation consists of the
creation of a new signal with parameter triple $\avec_{(i)}'$ independent of other
existing signals in the current model $\mathcal{M}_{k}$. The
one-to-one transformation in this case is very simply given by
$\textrm{t}_{k\mapsto k'}(\rvec)=\rvec =\avec_{(k+1)}'$
with Jacobian equal to 1.

Here, $q$ is the three-dimensional pdf from which we draw
proposals for the additional signal. We  use independent uniform
distributions with frequency range $0\leq f\leq 0.5$ and
amplitude range $(A_i^2+B_i^2)^{1/2}<l_A$ where $l_A$ is the
radius for the two amplitudes of the signal  in polar coordinates.
Again, the radii of the uniform proposal densities have to be tuned to
achieve an optimal acceptance rate.

\subsection{The delayed rejection method for parameter estimation}
For transitions within a model $\mathcal{M}_m$, classical MCMC
methods can be applied. Here, however, we use an adaptive MCMC
technique, the delayed rejection method
~\cite{Mira98,TierneyMira99,GreenMira01} that we have successfully
applied to estimate parameters of pulsars \cite{pul2}. The idea
behind the delayed rejection method is that persistent rejection
indicates that locally, the proposal distribution is badly
calibrated to the target. Therefore, the MH algorithm is modified
so that on rejection, a second attempt to move is made with a
proposal distribution that depends on the previously rejected
state. In this context, when a proposed MH move is rejected from a
bold normal distribution with large variance a second candidate
can be proposed with a timid proposal distribution for sampling
the parameters for the individual sinusoids. Hence, the main
objective of the first stage is a coarse scan of the parameter
space and therefore we choose the variances of the parameters
about one order of magnitude smaller than the prior ranges of the
corresponding parameters. Once a  mode is found, we aim to draw
representative samples in the second stage.

The precision of the frequency in a single-frequency model depends
on the amplitude, the variance $\sigma^2$ of the noise, and the
number of samples $N$ of the data set
\cite{Bretthorst88,Jaynes87}. The precision of the frequency has
been derived in \cite{Bretthorst88} by  a Gaussian approximation
to the posterior pdf of the frequency and calculation of its
standard deviation, given by $\sigma_{f}''=(2\pi)^{-1}[48 \sigma^2
/ N^3 (A^2+B^2)]^{1/2}$. We therefore choose proposals with this
standard deviation.

\subsection{Starting values}

The starting values of a Markov chain are crucial for the length
of the burn-in period, i.e.\ the time needed for the chain to achieve
 convergence to the real posterior distribution.
We perform a Fast
Fourier Transformation (FFT) prior to the simulation and use corresponding
estimates as starting values. Arthur Schuster introduced the
periodogram \cite{schuster05}
\begin{equation}
C(f)=\frac{1}{N} \left[ R(f)^2+I(f)^2 \right] \label{periodogram}
\end{equation}
where $R(f)=\sum_{j=1}^N{d_j \cos \left( 2 \pi f t_j\right)}$ and
$I(f)=\sum_{j=1}^N{d_j \sin \left( 2 \pi f t_j\right)}$ are the
real and imaginary parts from the sums of the discrete Fourier
transformation. As a starting value for $m$, we use the number
$m_0$ of local maxima in the periodogram that exceed a certain
noise level (lower than the expected one). We use the frequencies
corresponding to the local maxima in the periodogram, $f_{0,i}$,
as starting values for $f_{0,i}^{(m_0)}$ and $A_{0,i}=2
R(f_{0,i})/N$, $B_{0,i}=2 I(f_{0,i})/N$ as starting values for
$A_i^{(m_0)}$ and $B_i^{(m_0)}$, respectively.

\subsection{Identifying the sinusoids}
\label{classify} Although the RJMCMC offers great ease in model
selection, we still encounter the {\em label-switching} problem, a
general problem due to invariance of the likelihood under
relabelling that has been extensively discussed in the context of
mixture models \cite{Celeux00}.

The sinusoids that are contained in the model $\hat{m}$ with the highest posterior probability of $m$ are permutations of
$\hat{m}$ coexistent sinusoids out of a number of sinusoids that we do not know
but can estimate by the upper limit $m_{max}$ of the marginal posterior of $m$.
Therefore, the parameter vector sampled in each iteration of the
Markov chain (corresponding to model $\hat{m}$) is a permutation
of $\hat{m}$ parameter triples determining $\hat{m}$ out of $m_{max}$ sinusoids. The problem
is to determine which parameter triple belongs to which sinusoid.

The parameter that contributes significantly to identifying a
sinusoid is its frequency. We thus calculate the marginal
posterior of the frequency and obtain the $m_\textrm{max}$ strongest
peaks together with their frequency ranges by finding the threshold that
separates those peaks. It is still possible that individual peaks
contain more than one sinusoid or even none. This can be assessed by
a histogram simular to that in Fig.~\ref{fig_2} but restricted to the
 frequency range under consideration.
To separate more than one present sinusoids, we then
consider the two amplitudes and apply an agglomerative
hierarchical cluster analysis that involves all three parameters.
We use a modified Ward technique \cite{Ward63} that minimizes the
within group variance using a normalized Euclidean distance
between the parameters  by adjusting the frequency range to the
much larger range of the amplitudes.

\section{Results}
\label{results} 
\begin{figure}[hbt]
\centerline{\includegraphics[width=13cm]{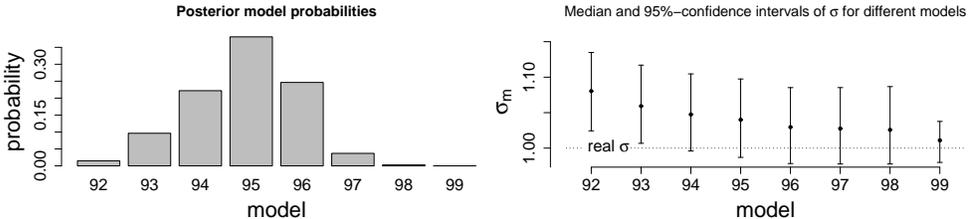} }
\caption{Posterior distribution of model number $m$ with 5\%,
50\%, and 95\% posterior confidence intervals of the corresponding
noise levels.} \label{fig_2}
\end{figure}

We created an artificial data set of 1\,000
samples from $m=100$ sinusoids. The sinusoids were randomly chosen
with maximum amplitudes $1$ and the noise standard deviation
was $\sigma=1$. We chose a uniform prior for $m$ on $\{0,1,2,
\ldots, M=60\,000\}$, and set $A_{\rm max}=B_{\rm max}=5$.  The Markov chain
ran for $10^8$ iterations and was thinned by storing every
$1\,000$th iteration. The first $5 \times 10^6$ iterations where
considered as burn-in and discarded. The MCMC simulation was
implemented in C on a 2.8\,GHz Intel P4 PC and took about 43 hours
to run. Fig.~\ref{fig_2} gives the histogram of the posterior model
probabilities obtained by the reversible jump algorithm. As each
model $\mathcal{M}_m$ is characterized by a different noise level
$\sigma_m$, we have also plotted the posterior distributions of
the noise standard deviations for increasing model order.
Note that $\sigma_m$ decreases with higher model order $m$ since a
model comprising more sinusoids accounts for more noise. Here, we
choose model $m=95$ corresponding to the posterior mode of $m$ as
the best fitting model.

\begin{figure}[hbt]
\centerline{\includegraphics[width=13cm]{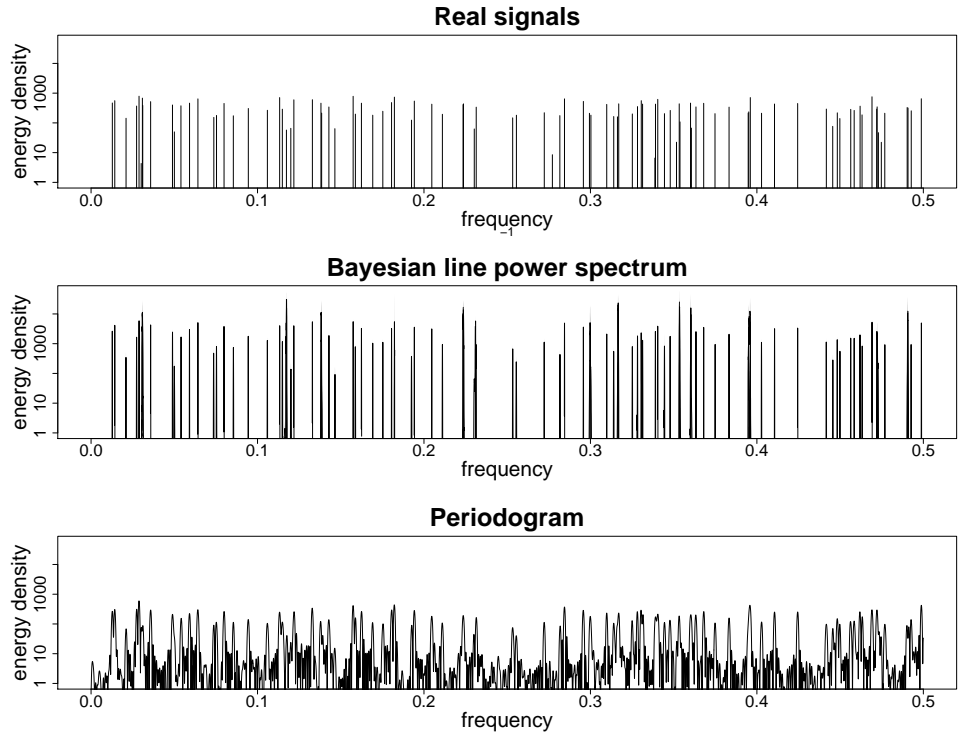} }
\caption{Comparison of real signals, classical Schuster
periodogram, and Bayesian spectrum. Note that the total energy of
the three spectra is similar but the lines have different
accuracies. Therefore the energy of a single line is more
concentrated due to the better accuracy of the Bayesian spectrum
estimate resulting in higher peaks. The theoretical spectrum
consists of delta functions for each sinusoid which would yield
infinite peaks. Therefore, an energy
contribution of $\frac{N}{2}(A_i^2+B_i^2)$ for each sinusoid with
frequency $f_i$ is shown.} \label{fig_3}
\end{figure}

We used all MCMC samples corresponding to model ${\cal M}_{95}$.
For ease of notation, we denote the parameter vector
of model ${\cal M}_{95}$ by
$(A_1,B_1,f_1,\ldots,A_{95},B_{95},f_{95})$. The complete line
power spectrum density can be estimated by the product of the
conditional expectation of the energy \linebreak $E(A_i^2+B_i^2|\dvec,m,f_i)$
of each sinusoid $i$ given its frequency $f_i$, and the posterior
pdf of $f_i$ given the data, $p(f_i|\dvec)$. One of the advantages
of the Bayesian spectrum analysis is the possibility to calculate
confidence areas for the spectrum. Therefore we group our MCMC
samples and calculate posterior confidence intervals for each
frequency bin. A sufficient width for the bins can be assessed by
the frequency accuracy $\sigma_{f}=(2\pi \cdot \textrm{snr})^{-1}
(48 / N^3)^{1/2}$ given by \cite{Bretthorst88}, where `snr' is the
signal-to-noise ratio. In our example, the choice of 30\,000 bins
is sufficient to resolve sinusoids with an snr of about $2$.
Fig.~\ref{fig_3} shows the real signals, the Bayes spectrum and
the classical Schuster periodogram mentioned in
Eq.~\ref{periodogram}.

The plot for the real signals displays an individual energy
contribution for each sinusoid $i$ of $(A_i^2+B_i^2) N/2$.
Normally a theoretical spectrum would consist of delta functions
with infinitely large energy peaks since the energy contribution
is concentrated on an interval of infinitely small width.
Therefore we just plotted the energy contribution on the energy
scale that yields a similar scaling as obtained by the
periodogram.

\begin{figure}[hbt]
\centerline{ \includegraphics[width=13cm]{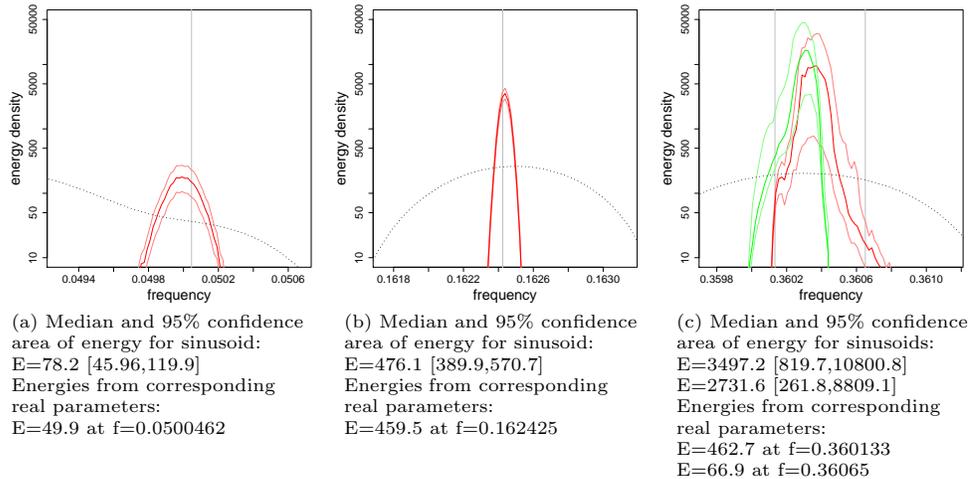} }
\begin{tabular}{p{4cm}p{4cm}p{4cm}p{2cm}} \scriptsize \raggedright
(a) Median and 95\% confidence area of energy for sinusoid: \linebreak
E=78.2 [45.96,119.9] \linebreak
Energies from corresponding real parameters: \linebreak
E=49.9 at f=0.0500462
& \scriptsize \raggedright
(b) Median and 95\% confidence area of energy for sinusoid: \linebreak
E=476.1 [389.9,570.7] \linebreak
Energies from corresponding real parameters: \linebreak
E=459.5 at f=0.162425
& \scriptsize \raggedright
(c) Median and 95\% confidence area of energy for sinusoids: \linebreak
E=3497.2 [819.7,10800.8] \linebreak
E=2731.6 [261.8,8809.1] \linebreak
Energies from corresponding real parameters: \linebreak
E=462.7 at f=0.360133 \linebreak
E=66.9 at f=0.36065
&
\end{tabular}
\caption{Three magnified areas for a comparison of the classical
periodogram (dotted line) and the 95\% confidence area (area
within bright colored lines) of the Bayesian spectrum. The darker
colors within the 95\% confidence area indicate the median of the
energy. The vertical lines show the delta functions of the
theoretical spectrum. Its energies are given above.} \label{fig_4}
\end{figure}

In order to be able to display 95\% confidence areas of the
spectrum we present three magnified areas in Fig.~\ref{fig_4}. In
plot (a) we see a sinusoid with rather small energy. The accuracy
of the frequency estimation is worse compared with the sinusoid of
graph (b) which has a significantly larger energy. The third graph
shows two very close sinusoids. The frequency estimation is very
inaccurate due to the interference of the two signals. This is
consistent with theoretical results by \cite{Bretthorst88}. The
interference of the two close signals is due to a phase shift of
$175^\circ$. The interference and hence the frequency estimation
depends upon whether the sinusoids are orthogonal
\cite{Bretthorst88} or not. Nevertheless, we are able to identify
the existence of two signals while the periodogram only reveals
the existence of one.

The estimates of the amplitudes, however, always show huge
values and confidence intervals for sinusoids close in frequency.
The huge energies are merely restricted by the choice of priors
for the amplitudes. The reason for this is due to the possible
combinations to express a sum of sinusoids when the observation
time is insufficiently long with respect to the distance in
frequency. In this case we can not make accurate statements about
the amplitudes and hence energies of both sinusoids.

If we take a look at the single peak of the periodogram the energy
that is considered is subject to the data from a discrete and
finite observation time, given by
$\sum_{j=1}^{N}{\mathrm{f}_m(t_j,\avec_m)}$. This, however does
not reflect the energy contribution of the real signals. The
Bayesian estimates of the amplitudes are honest by yielding large
confidence areas for the energies of sinusoids close in
frequencies but in return small confidence intervals for isolated
sinusoids.

\section{Discussion}
\label{disc} We have presented a Bayesian approach to identifying
a large number of unknown periodic signals embedded in noisy data.
A reversible jump MCMC technique can be used to estimate the
number of signals present in the data, their parameters, and the
noise level. This approach allows for simultaneous detection and
parameter estimation, and does not require a stopping criterion
for determining the number of signals. The MCMC method compares
favorably with classical spectral techniques.

Our motivation for this research is to address the difficulty that
LISA will ultimately encounter in having too many signals present.
LISA may \emph{see} 100\,000 signals from binary systems in the 1
mHz to 5 mHz band. We see our work as a new method that could help
LISA to identify and characterize these signals. The work here is
a simplified problem, one that neglects the time evolution of the
signal and modulation due to LISA's orbit. The next step is to
deal with these more complicated signals, and to develop a
realistic strategy for applying our MCMC methods to more realistic
LISA data. We believe that MCMC methods, like those presented
here, provide a practical and highly effective method of
identifying and characterizing the large number of signals that
will exist in the LISA data.

\verb''\ack This work was supported by National Science Foundation
grants PHY-0071327 and PHY-0244357, the Royal Society of New
Zealand Marsden fund award UOA204, Universities UK, and the
University of Glasgow.


\Bibliography{9}
\bibitem{LISA} K. Danzmann and A. R\"udiger, Classical and Quantum Gravity {\bf 20}, S1 (2003)
\bibitem{cutler} L. Barak and C. Cutler, Physical Review D {\bf 70}, 122002 (2004)
\bibitem{bena} M.J. Benacquista, J. DeGoes, D. Lunder, Classical and Quantum Gravity {\bf 21}, S509-S514 (2004)
\bibitem{cornish} J. Crowder and N.J. Cornish, Physical Review D {\bf 70}, 082004 (2004)
\bibitem{nele01} G. Nelemans, L.R. Yungleson and S.F. Protegies Zwart, Astronomy and Astrophysics, {\bf 375}, 890 (2001)
\bibitem{gilks96}  W.R. Gilks and S. Richardson and D.J. Spiegelhalter, {\it
Markov Chain Monte Carlo in Practice}, Chapman and Hall, London (1996).
\bibitem{metr53}  N. Metropolis, A.W. Rosenbluth, M.N. Rosenbluth, A.H.
Teller, E. Teller, Journal of Chemistry and Physics {\bf 21}, 1087 (1953).
\bibitem{hastings70}  W.K. Hastings, Biometrika {\bf 57}, 97 (1970).
\bibitem{insp}  N. Christensen, R. Meyer and A. Libson,
Class. Quant. Grav. {\bf 21}, 317 (2004).
\bibitem{pul1} N. Christensen, R.J. Dupuis, G. Woan and R. Meyer, Phys. Rev D {\bf 70}, 022001 (2004).
\bibitem{pul2} R. Umst\"atter, R. Meyer, R.J. Dupuis, J. Veitch, G. Woan and N. Christensen, Classical and Quantum Gravity {\bf 21}, S1655 (2004).
\bibitem{LISAMCMC1} R Umst\"atter, N Christensen, M Hendry, R Meyer, V Simha, J Veitch, S Vigeland and Graham Woan, {\it Bayesian Modeling of source confusion in LISA data}, pre-print.
\bibitem{Jaynes03}  E.T. Jaynes and G. L. Bretthorst , {\it
Probability Theory : The Logic of Science}, (Cambridge University Press) (2003).
\bibitem{cornish2} N.J. Cornish and S. L. Larson Physical Review D {\bf 67} 103001 (2003)
\bibitem{Rich} S. Richardson and P.J. Green, J.R.Statist. Soc. B {\bf 59}, 731 (1997).
\bibitem{Andrieu} C. Andrieu, A. Doucet, IEEE Trans. Sig. Proc. {\bf 47}, 2667 (1999).
\bibitem{Bretthorst88} G. L. Bretthorst, \emph{Bayesian Spectrum Analysis and Parameter Estimation}, (Springer Lecture Notes in Statistics \#48), (1988).
\bibitem{Green95} P.J. Green, Biometrika {\bf 82}, 711 (1995)
\bibitem{Green03} P.J. Green, {\it In Highly Structured Stochastic Systems}
(Green, P.J., Hjort, N.L., Richardson, S., editors), Oxford University Press (2003)
\bibitem{Mira98} A. Mira {\it Ordering, Slicing and Splitting Monte Carlo Markov chain}
PhD thesis, University of Minnesota (1998)
\bibitem{TierneyMira99} L. Tierney and A. Mira Statistics in Medicine {\bf 18} 2507 (1999)
\bibitem{GreenMira01} P. J. Green and A. Mira, Biometrika {\bf 88}, 1035 (2001)
\bibitem{Jaynes87} E. T. Jaynes, \emph{Bayesian Spectrum Analysis and Chirp Analysis}, Maximum Entropy and Bayesian Spectral Analysis and Estimation Problems, C. Ray Smith, and G.J. Erickson, ed., D. Reidel, Dordrecht-Holland, 1-37 (1987)
\bibitem{schuster05} A. Schuster, Proceedings of the Royal Society of London, {\bf 77} 136 (1905)
\bibitem{Celeux00} G. Celeux, M. Hurn, C. P. Robert, Journal of the American Statistical Association {\bf 95} 957-970 (2000)
\bibitem{Ward63} J.H. Ward, Journal of the American Statistical Association, {\bf 58}, 236-244 (1963)
\endbib
\end{document}